\documentclass{ws-procs}

\newcommand\rf[1]{(\ref{eq:#1})}
\newcommand\lab[1]{\label{eq:#1}}
\newcommand\nonu{\nonumber}
\newcommand\br{\begin{eqnarray}}
\newcommand\er{\end{eqnarray}}
\newcommand\be{\begin{equation}}
\newcommand\ee{\end{equation}}

\newcommand\lb{\lbrack}
\newcommand\rb{\rbrack}

\newcommand\llb{\left\lbrack}
\newcommand\rrb{\right\rbrack}

\renewcommand\({\left(}
\renewcommand\){\right)}
\renewcommand\v{\vert}                     

\newcommand\bc{\begin{center}}
\newcommand\ec{\end{center}}

\newcommand\Tr{\mathop{\mathrm Tr}}                  
\newcommand\partder[2]{\frac{{\partial {#1}}}{{\partial {#2}}}}

\renewcommand\a{\alpha}
\renewcommand\b{\beta}

\renewcommand\d{\delta}
\newcommand\D{\Delta}
\newcommand\eps{\epsilon}
\newcommand\vareps{\varepsilon}
\newcommand\g{\gamma}
\newcommand\G{\Gamma}
\newcommand\grad{\nabla}
\newcommand\h{\frac{1}{2}}
\renewcommand\k{\kappa}

\renewcommand\L{\Lambda}
\newcommand\m{\mu}
\newcommand\n{\nu}
\renewcommand\o{\over}

\renewcommand\O{\Omega}
\newcommand\p{\phi}
\newcommand\vp{\varphi}
\renewcommand\P{\Phi}
\newcommand\pa{\partial}

\newcommand\pr{\prime}

\newcommand\s{\sigma}

\renewcommand\t{\tau}

\newcommand\z{\zeta}

\newcommand\wti{\widetilde}

\newcommand\cA{{\mathcal A}}

\newcommand\cD{{\mathcal D}}

\newcommand\cF{{\mathcal F}}

\newcommand\cL{{\mathcal L}}

\newcommand\cP{{\mathcal P}}

\newcommand\cT{{\mathcal T}}

\newcommand{\ct}[1]{\cite{#1}}
\newcommand{\bib}[1]{\bibitem{#1}}
\newcommand\PRD[3]{(#2), \textsl{Phys. Rev.} \textbf{D#1} #3}

\newcommand\PLB[3]{(#2), \textsl{Phys. Lett.} \textbf{#1B} #3}
\newcommand\CQG[3]{(#2), \textsl{Class. Quantum Grav.} \textbf{#1} #3}

\newcommand\PR[3]{(#2), \textsl{Phys. Reports} \textbf{#1} #3}

\begin{document}
\title{Impact of Dynamical Tensions in Modified String and Brane Theories}
\author{E.I. Guendelman and A. Kaganovich}\address{\small\it 
Department of Physics,\\[-1.mm]
\small\it Ben-Gurion University, Beer-Sheva, Israel  \\[-1.mm]
\small\it email: guendel@bgumail.bgu.ac.il, alexk@bgumail.bgu.ac.il}  
\author{E. Nissimov and S. Pacheva}\address{\small\it Institute for Nuclear Research and Nuclear Energy,\\[-1.mm]
\small\it Bulgarian Academy of Sciences, Sofia, Bulgaria  \\[-1.mm]
\small\it email: nissimov@inrne.bas.bg, svetlana@inrne.bas.bg}  
\maketitle\abstracts{First we briefly outline the general construction of
$Dp$-brane models with {\em dynamical} tensions. We then proceed to a more 
detailed discussion of a modified string model where the string
tension is related to the potential of (an external) world-sheet electric
current. We show that cancellation of the pertinent conformal anomaly on 
the quantum level requires the dynamical string tension to be a square of 
a free massless world-sheet scalar field.}
\section{Main Motivation}
Dirichlet $p$-branes ($Dp$-branes) \ct{Dp-brane-orig} are $p+1$-dimensional 
extended objects in space-time which carry the end points of fundamental 
open strings. Their crucial relevance in modern string theory is due
to several basic properties of theirs such as providing
explicit realization of non-perturbative string dualities, microscopic 
description of black-hole physics, gauge theory/gravity correspondence, 
large-radius compactifications of extra dimensions, brane-world scenarios in 
particle phenomenology, \textit{etc.} . For a background on string and 
brane theories, see refs.\ct{brane-string-rev}. 

In an independent recent development two of us have proposed a broad class of 
new models involving Gravity called \textsl{Two-Measure Gravitational Models}
\ct{TMT}, whose actions are typically of the form:
\be
S = \int d^D x\, \P (\vp)\, L_1 + \int d^D x\,\sqrt{-g}\, L_2 \; ,
\lab{TMT-a}
\ee
\be
L_{1,2} = e^{\frac{\a \phi}{M_P}} \Bigl\lb - {1\o \k} R(g,\G ) -
\h g^{\m\n}\pa_\m \phi \pa_\n \phi
+ \bigl(\mathrm{Higgs}\bigr) + \bigl(\mathrm{fermions}\bigr)\Bigr\rb \; .
\lab{TMT-b}
\ee
with the standard notations: $R(g,\G )$ is the scalar curvature in the 
first-order formalism (\textsl{i.e.}, the connection $\G$ is independent of 
the metric $g_{\m\n}$), $\phi$ is the dilaton field, $M_P$ is the Planck mass,
\textsl{etc.}. The main new ingredient appears in the first term of \rf{TMT-a}
-- it is an alternative non-Riemannian (i.e. independent of the metric 
$g_{\m\n}$) generally-covariant integration measure density $\P (\vp)$ built 
up in terms of additional auxiliary scalar fields $\vp^i$ ($i=1,\ldots ,D$ 
where $D$ is the space-time dimension) :
\be
\P (\vp) \equiv {1\o {D!}}\vareps^{\m_1 \ldots \m_D}
\vareps_{i_1 \ldots i_D} \pa_{\m_1} \vp^{i_1} \ldots \pa_{\m_D} \vp^{i_D} \; .
\lab{D-measure-def}
\ee
Although naively the additional ``measure-density'' scalars $\vp^i$ appear in 
\rf{TMT-a} as pure-gauge degrees of freedom (due to the invariance under arbitrary
diffeomorphisms in the $\vp^i$-target space), there is still a remnant -- the so
called ``geometric'' field $\chi (x) \equiv \frac{\P (\vp)}{\sqrt{-g}}$, which 
remains an additional dynamical degree of freedom beyond the standard physical 
degrees of freedom characteristic to the ordinary gravity models with the standard
Riemannian-metric integration measure. The most important property of the 
``geometric'' field $\chi (x)$ is that its dynamics is determined solely through the
matter fields locally (\textsl{i.e.}, without gravitational interaction).
The latter turns out to have a significant impact on the physical properties
of the two-measure gravity models which allows them to address various basic
problems in cosmology and particle physics phenomenology and provide
physically plausible solutions, for instance: (i) the issue of scale invariance
and its dynamical breakdown, \textsl{i.e.}, spontaneous generation of
dimensionfull fundamental scales; (ii) cosmological constant problem;  
(iii) geometric origin of fermionic families. In the very recent papers 
\ct{conf-inv-bworld} it has been demonstrated that two-measure gravity theories
are of significant interest in the context of modern brane-world scenarios,
namely, a new conformally invariant brane-world model in $D\! =\! 6$ 
without (bulk) cosmological constant fine tuning has been constructed there.

Subsequently, the idea of employing alternative non-Riemannian
reparametrization-covariant integration measures was applied in the context
of strings and branes theories \ct{mstring-1,kopa}. A common
basic property of these modified string/brane models is that the ratio of
both integration measure densities (the alternative versus the standard
Riemannian) becomes a dynamical string/brane tension -- an additional
dynamical degree of freedom beyond the original string/brane degrees of
freedom. In particular, in ref.\ct{kopa} we systematically constructed a new
class of modified $Dp$-brane models with dynamical brane tension -- this
construction is briefly reviewed in Section 2. We have shown in 
\ct{mstring-1,kopa}, that the dynamical nature of the string/brane tension
leads to some new interesting physical effects such as simple mechanisms of
confinement of ``color'' point-like charges (in the string case) and of charged
lower-dimensional sub-branes (in the $Dp$-brane case).

In Section 3 below we study in some detail the quantum properties of a
modified string model where the dynamical tension becomes related
to the potential of an external electric charge current on the string
world-sheet. We explicitly show that quantum consistency, \textsl{i.e.},
cancellation of the pertinent conformal anomaly, apart from the well-known
restriction on $D$ ($D\! =\! 26$ in the simplest bosonic case) implies that
the dynamical string tension must be a square of a free massless world-sheet
scalar field.

\section{$Dp$-Branes with Dynamical Tension}
First, let us recall the standard formulation of $Dp$-branes given in terms of 
the Dirac-Born-Infeld (DBI) action (see \textsl{e.g.} third 
ref.\ct{brane-string-rev}) :
\be
S_{DBI} = -T\int d^{p+1}\s\,\llb e^{-\a U}\sqrt{-\det\vert\vert G_{ab} - \cF_{ab} \vert\vert}
+ \ldots \rrb
\; ,
\lab{DBI-action}
\ee
with the following short-hand notations:
\be
G_{ab} \equiv \pa_a X^\m \pa_b X^\n G_{\m\n}(X) \quad ,\quad
\cF_{ab} \equiv \pa_a X^\m \pa_b X^\n B_{\m\n}(X) - F_{ab}(A)\; .
\lab{G-F-notations}
\ee
Here $G_{\m\n}(X)$,  $U(X)$, and $B_{\m\n}(X)$ and are the 
background metric, the dilaton, and the $2$-form Neveu-Schwarz, respectively, 
jwhereas $F_{ab}(A) = \pa_a A_b - \pa_b A_b$ is the field-strength of the 
Abelian world-volume gauge field $A_a$. The dots in \rf{DBI-action} indicate
coupling to the $(p+1)$-form Ramond-Ramond background gauge field which is 
omitted for simplicity. All world-volume indices take values 
$a,b=0,1,\ldots ,p$ and $\vareps^{a_1 \ldots a_{p+1}}$ is the 
$(p+1)$-dimensional totally antisymmetric tensor ($\vareps^{0 1\ldots p} = 1$).

Similarly to the gravity case \rf{TMT-a}--\rf{D-measure-def} we now introduce 
a modified world-volume integration measure density in terms of $p+1$ auxiliary
scalar fields $\vp^i$ ($i=1,\ldots ,p+1$) :
\be
\P (\vp) \equiv \frac{1}{(p+1)!} \vareps_{i_1\ldots i_{p+1}} 
\vareps^{a_1\ldots a_{p+1}} \pa_{a_1} \vp^{i_1}\ldots \pa_{a_{p+1}} 
\vp^{i_{p+1}} \; ,
\lab{brane-measure}
\ee
and use it to construct the following new $p$-brane-type action :
\be
S = - \int d^{p+1}\s\, \P (\vp) \Bigl\lb e^{-\b U} \h\z^{ab}\( G_{ba} - \cF_{ba}\)
+ {1\o \sqrt{-\z}} \O (\cA)\Bigr\rb + \int d^{p+1}\s\,\cL (\cA)  \; .
\lab{mDBI-action-0}
\ee
Here apart from \rf{G-F-notations} the following new notations
are used. The $(p+1)\times (p+1)$ matrix $\z_{ab}$ of auxiliary variables is
an arbitrary world-volume $2$-tensor, $\z^{ab}$ denotes the corresponding
inverse matrix ($\z^{ac} \z_{cb} = \d^a_b$) and $\z \equiv \det\v\v\z_{ab}\v\v$. 
The term $\O (\cA)$ indicates a topological density given in terms of some 
additional auxiliary gauge fields $\cA^I$ living on the world-volume:
\be
\partder{\O}{\cA^I} - \pa_a \(\partder{\O}{\pa_a \cA^I}\) = 0 \;\;
\mathrm{identically} \quad ,\quad
\mathrm{i.e.}\;\; \d\O (\cA) = \pa_a \(\partder{\O}{\pa_a \cA^I} \d \cA^I\) \; . 
\lab{top-density-def}
\ee
Finally, $\cL (\cA)$ describes possible coupling of the auxiliary fields $A^I$ to
external ``currents'' on the brane world-volume.

The requirement for $\O (\cA)$ to be a topological density is dictated by the
requirement that the new brane action \rf{mDBI-action-0} (in the 
absence of the last gauge-coupling term $\int d^{p+1}\s\,\cL (\cA)$) reproduces 
the standard $Dp$-brane equations of motion resulting from the DBI action
\rf{DBI-action} apart from the fact that the $Dp$-brane tension 
$T \equiv \P (\vp)/\sqrt{-\z}$ becomes now an {\em additional dynamical degree
of freedom} (note that no \textit{ad hoc} dimensionfull tension factor $T$ has
been introduced in \rf{mDBI-action-0}). 
Let us particularly stress that the modified-measure
brane model \rf{mDBI-action-0} naturally requires (through the necessity to
introduce topological density $\O (\cA)$) the existence on the
world-volume of an additional (higher-rank tensor) gauge field $\cA$ apart
from the standard world-volume Abelian vector gauge field $A_a$.

Splitting the auxiliary tensor variable $\z^{ab} = \g^{ab} + \z^{[ab]}$ into
symmetric and anti-symmetric parts and setting $\z^{[ab]}=0$, the action
\rf{mDBI-action-0} reduces to the action of the modified-measure model of ordinary
$p$-branes \ct{mstring-1} with Neveu-Schwarz field $B_{\m\n}$ and
world-volume gauge field $A_a$ disappearing and $\g_{ab}$ assuming the role of 
world-volume Riemannian metric.

The most obvious example of a topological density $\O (\cA)$ for the
additional auxiliary world-volume gauge fields in \rf{mDBI-action-0} is:
\be
\O (\cA) = -\frac{\vareps^{a_1\ldots a_{p+1}}}{p+1} F_{a_1\ldots a_{p+1}} (\cA)
\quad ,\quad
F_{a_1\ldots a_{p+1}} (\cA) = (p+1)\pa_{\lb a_1} \cA_{a_2\ldots a_{p+1}\rb} \; ,
\lab{top-density-p-rank}
\ee
where $\cA_{a_1 \ldots a_p}$ denotes rank $p$~ antisymmetric tensor (Abelian) 
gauge field on the world-volume. More generally we can have 
(for $p+1\! =\! rs$) : 
\be
\O (\cA) = {1\o {rs}}\vareps^{a_{11}\ldots a_{1r} \ldots a_{s1}\ldots a_{sr}} 
F_{a_{11}\ldots a_{1r}}(\cA) \ldots F_{a_{s1}\ldots a_{sr}}(\cA)
\lab{top-density-rs}
\ee
with rank $r-1$ (smaller than $p$) auxiliary world-volume gauge fields.

We may also employ {\em non-Abelian}~ auxiliary world-volume gauge fields 
$\cA_a$. For instance, when $p+1\! =\! 2q$ :
\be
\O (\cA) = 
{1\o {2q}}\vareps^{a_1 b_1\ldots a_q b_q} 
\Tr\( F_{a_1 b_1}(\cA)\ldots F_{a_q b_q}(\cA) \) \; ,
\lab{top-density-NA-q}
\ee
where $F_{ab}(\cA) = \pa_a\cA_b - \pa_b\cA_a + i \bigl\lb \cA_a,\,\cA_b\bigr\rb$.

The equations of motion w.r.t. $\vp^i$ and $\z^{ab}$ corresponding to the 
modified-measure brane action \rf{mDBI-action-0} read: 
\be
e^{-\b U} \h \z^{ab} \( G_{ba} - \cF_{ba}\) + {1\o \sqrt{-\z}} \O (\cA) = 
M \equiv \mathrm{const} \; ,
\lab{brane-vp-eqs}
\ee
\be
e^{-\b U} \( G_{ab} - \cF_{ab}\) + \z_{ab} \frac{1}{\sqrt{-\z}} \O (\cA) = 0 \; .
\lab{z-eqs}
\ee
Both Eqs.\rf{brane-vp-eqs}--\rf{z-eqs} imply:
\be
\z^{ab} \( G_{ba} - \cF_{ba}\) = 2M \frac{p+1}{p-1} e^{\b U} \quad ,\quad
\frac{1}{\sqrt{-\z}} \O (\cA) = - \frac{2M}{p-1}
\lab{brane-trace-eqs}
\ee
which when substituted in \rf{z-eqs} give:
\be
G_{ab} - \cF_{ab} = \frac{2M}{p-1} e^{\b U}\,\z_{ab}
\lab{G-F-z-eqs}
\ee

Next we consider the equations of motion 
w.r.t. auxiliary (gauge) fields $\cA^I$ :
\be
\pa_a \Bigl(\frac{\P (\vp)}{\sqrt{-\z}}\Bigr)\,\partder{\O}{\pa_a \cA^I} + j_I = 0
\; ,
\lab{brane-A-eqs}
\ee
where $j_I \equiv \partder{\cL}{\cA^I} - \pa_a \Bigl(\partder{\cL}{\pa_a \cA^I}\Bigr)$
is the corresponding ``current'' coupled to $\cA^I$. These are the equations
determining the dynamical brane tension $T \equiv \P (\vp)/\sqrt{-\z}$. 
In deriving Eq.\rf{brane-A-eqs} crucial use was made of the identity
\rf{top-density-def} satisfied by the topological density $\O (\cA)$.

In particular, in the absence of coupling of external world-volume currents
to the auxiliary (gauge) fields $\cA^I$ Eq.\rf{brane-A-eqs} imply:
\be
T \equiv \P (\vp)/\sqrt{-\z} = C \equiv \mathrm{const}
\lab{brane-tension-const}
\ee

Now, using Eqs.\rf{brane-vp-eqs} and \rf{G-F-z-eqs} it is straightforward to 
show that the modified brane action \rf{mDBI-action-0} with $\cL (\cA)=0$
classically reduces to the standard $Dp$-brane DBI-action \rf{DBI-action} :
\be
S^\pr_{DBI} = -T^\pr \int d^{p+1}\s\, e^{-\b^\pr U}
\sqrt{-\det\vert\vert G_{ab} - \cF_{ab} \vert\vert}  \; ,
\lab{DBI-action-a}
\ee
\be
T^\pr \equiv \h C (2M)^{-\frac{p-1}{2}} (p-1)^{\frac{p+1}{2}} \quad ,\quad 
\b^\pr \equiv \frac{p+1}{2}\b \; ,
\lab{brane-tension-const-a}
\ee
where, however, the $Dp$-brane tension $T^{\pr}$ is dynamically generated
according to \rf{brane-tension-const} and \rf{brane-tension-const-a}.

For a more detailed analysis of the properties of the modified-measure
$Dp$-brane models we refer to \ct{kopa}.

\section{Conformal Anomaly and Its Impact on the Dynamical String Tension}
Now we turn our attention to a special case of \rf{mDBI-action-0} for
$p\! =\! 1$, \textsl{i.e.}, a modified string model with dynamical tension:
\be
S = - \int d^2\s \,\P (\vp) \Bigl\lb \h \g^{ab} \pa_a X^{\m} \pa_b X_{\m} -
\frac{\vareps^{ab}}{2\sqrt{-\g}} F_{ab} (A) \Bigr\rb
+ \eps \int d^2\s \, A_a \vareps^{ab} \pa_b u  \; .
\lab{action-string-u}
\ee
Notice that the last term in Eq.\rf{action-string-u} can be rewritten
in the reparametrization-invariant form:
\be
\int d^2\s\,\sqrt{-\g}\, A_a J^a \quad ,\quad
J^a \equiv \eps \frac{\vareps^{ab}}{\sqrt{-\g}} \pa_b u
\; ,
\lab{action-string-u-last}
\ee
where $J^a$ indicates the general expression for a covariantly conserved 
world-sheet electric current. Let us stress that such coupling of string
degrees of freedom to an external world-sheet electric current is natural
only in the present context of modified-measure string models due to the
inevitable appearance of the auxiliary world-sheet gauge field $A_a$.
Let us also note that in \rf{action-string-u-last} the ordinary Riemannian
world-sheet integration measure density $\sqrt{-\g}$ is used unlike the 
modified one:
\be
\P (\vp)\! =\! \h \vareps_{ij}\vareps^{ab} \pa_a \vp^i \pa_b \vp^j
\quad ,\quad a,b =0,1 \;\; ,\;\; i,j =1,2
\lab{string-measure}
\ee
in the main action term in \rf{action-string-u}, much in the spirit of the
previously proposed two-measure gravity theories \ct{TMT} (cf. Eq.\rf{TMT-a}). 

Quantization of the modified string model \rf{action-string-u} within the
functional integral framework can be performed in the standard way based on
the canonical Hamiltonian formalism for constrained systems a'la Dirac. 
As already shown in the third ref.\ct{mstring-1}, the total canonical Hamiltonian
$H_T \equiv \sum_A \L_A \P_A$ is a linear combination of the following first-class 
constraints $\P_A$ (the letters $\pi$ and $\cP$ indicating the pertinent canonical
momenta) :
\be
\quad \pi_{\g^{ab}} = 0 \quad ,\quad
\cT_{\pm} \equiv \frac{1}{4} 
\Bigl(\frac{\cP}{E} \pm \pa_\s X \Bigr)^2 = 0
\; ,
\lab{standard-constr}
\ee
which are of the same form as in the ordinary bosonic string case modulo the
fact that now the string tension $T\equiv E$ is a dynamical degree of freedom
(see last Eq.\rf{A-constr} below), plus the new constraints:
\be
\pa_\s \vp^i \pi^{\vp}_i = 0  \quad ,\quad \frac{\pi^{\vp}_2}{\pa_\s \vp^1} = 0 
\; ;
\lab{phi-constr}
\ee
\be
\pi_{A_0} = 0 \quad , \quad \pa_\s (E + \eps u) = 0 \quad ,\;\; \mathrm{where} 
\;\; E \equiv \pi_{A_1} = \frac{\P (\vp)}{\sqrt{-\g}} \; .
\lab{A-constr}
\ee
Relations \rf{phi-constr} imply that the auxiliary measure-fields $\vp^i$
are pure-gauge degrees of freedom which decouple from the string dynamics.
The only remnant of the latter appears through $E$ -- the canonical momentum
of $A_1$ according to the last expression \rf{A-constr}, which together with
the second relation \rf{A-constr} tells us that $E$ has the physical meaning of
a world-sheet electric field-strength obeying the $D=2$ Gauss law.

To quantize the modified-measure string model \rf{action-string-u}
one starts with the standard Faddeev's functional integral:
\br
Z = \int \cD X \,\cD\cP \,\cD E \,\cD A_1 \cD \L_A \,\D_{\P\Pi} 
\,\d (\mathrm{conf.~gauge})
\nonu \\
\times \exp\Bigl\{ i\int d^2\s\,\Bigl\lb \cP_\m \pa_\t X^\m + E \pa_\t A_1 -
\sum_A \L_A \P_A \Bigr\rb\Bigr\}
\lab{Z-1}
\er
where $\P_A$ are the first-class constraints listed above 
\rf{standard-constr}--\rf{A-constr} and $\D_{\P\Pi}$ indicates the 
Faddeev-Popov ghost determinant associated with the conformal gauge-fixing 
condition. In \rf{Z-1} and in what follows we shall skip the insertion of
vertex operators for brevity. Now performing the Gaussian integrations over 
the canonical momenta we arrive at the following reparametrization-invariant 
expression:
\br
Z = \int \cD X\,\cD\g_{ab}\,\cD E\,\cD A_a\,\D_{\P\Pi} \,\d (\mathrm{conf.~gauge})
\nonu \\
\times
\exp\Bigl\{ i\int d^2\s\,\Bigl\lb - E\h \sqrt{-\g}\g^{ab} \pa_a X^{\m}\pa_b X_{\m}
+ \h (E+\eps u) \vareps^{ab} F_{ab}(A)\Bigr\rb \Bigr\}
\lab{Z-2}
\er
Integration over the auxiliary gauge field $A_a$ yields functional
delta-function $\d \(\vareps^{ab}\pa_b (E+\eps u)\)$ 
which in turn reduces the functional integration over $E$ to an
ordinary integration over the overall world-sheet constant $C$ :
\br
Z = \int dC \,\cD X \,\cD\g_{ab} \,\D_{\P\Pi} \,\d (\mathrm{conf.~gauge})
\nonu \\
\times \exp\Bigl\{ -i\h\int d^2\s\, 
(C-\eps u)\sqrt{-\g }\g^{ab} \pa_a X^{\m}\pa_b X_{\m} \Bigr\}
\lab{Z-3}
\er
Note that $T \equiv C-\eps u$ is the dynamical string tension.
Thus, integration over string coordinates $X^\m$ amounts to computation of
the determinant of the modified Dalembertian (or Laplace-Beltrami upon
Euclidean rotation) operator:
\be
- {1\o \sqrt{-\g}} \pa_a \bigl( v^2 \sqrt{-\g} \g^{ab} \pa_b\bigr)
\lab{box-v}
\ee
where for later convenience we have introduced the notation $v$ for the
square-root of the dynamical tension: $v^2 \equiv C - \eps u$. 
Due to the presence of the latter there is an additional contribution to the
well-known conformal anomaly (see \textsl{e.g.} \ct{vassil} and references 
therein where the ``dilaton''-like notation $v = e^{-\p}$ is employed) :
\br
Z = \int dC \,\cD \g_{ab}\,\d (\mathrm{conf.~gauge})
\exp\Bigl\{ i\int d^2\s\,\sqrt{-\g}\Bigl\lb \frac{D-26}{96\pi}\, R\Box^{-1}R
\nonu \\
+ {D\o {8\pi}}\,\frac{\Box v}{v}\, \Box^{-1}R 
- b{D\o{4\pi}} \g^{ab}(\grad_a \ln v) (\grad_b \ln v) \Bigr\rb\Bigr\}
\lab{Z-anomal}
\er
Here $R$ denotes the usual scalar curvature for the intrinsic world-sheet metric
$\g_{ab}$ and $\Box \equiv {1\o \sqrt{-\g}}\pa_a\bigl(\sqrt{-\g}\g^{ab}\pa_b\bigr)$ 
is the ordinary covariant Dalembertian. The last non-anomalous term in 
\rf{Z-anomal} is determined up to a regularization-dependent constant $b$. 
Therefore, absence of conformal anomaly  implies in the present
case, apart from the usual critical value for the space-time dimension,
an additional condition on the square-root of the dynamical string tension:
\be
\Box v = 0 \quad, \quad v^2 \equiv T = C-\eps u \; ,
\lab{free-massless}
\ee
\textsl{i.e.}, the dynamical string tension must be a square of a free
massless world-sheet scalar field. Note that this condition is a dynamical
constraint on the external world-sheet electric current \rf{action-string-u-last}
coupled to the modified-measure string.

Going back to the string action in \rf{Z-3} and taking into account relations
\rf{free-massless} we can rewrite it in the following simple form:
\be
-\h \int d^2\s\,\sqrt{-\g}\g^{ab}\pa_a\bigl( vX^\m\bigr)\pa_b\bigl( vX_\m\bigr)
\lab{action-string-v}
\ee
where we used the following identity for the modified Dalembertian
operator \rf{box-v} :
\be
{1\o \sqrt{-\g}} \pa_a \bigl( v^2 \sqrt{-\g} \g^{ab} \pa_b\bigr)
\bigl(\cdot\bigr) = v\,\Box\bigl( v \cdot \bigr) - \bigl(\Box v\bigr)\, v = 
v\,\Box\bigl( v \cdot\bigr)
\lab{box-v-id}
\ee
due to the free masslessness of the square-root tension $v$. Therefore, in 
the quantized modified-measure string model \rf{action-string-u} it is 
the dynamically rescaled fields $\wti{X}^\m = v\,X^\m$ which describe free 
wave mode propagation along the string rather than the usual string coordinates
$X^\m$.

\section{Conclusions}
Replacing the standard Riemannian world-sheet/world-volume integration
measure density with an metric-independent reparametrization-invariant one 
\rf{brane-measure} in the Lagrangian formulation of string and brane models
has significant impact on the string/brane dynamics. Consistency of dynamics
requires the introduction of additional auxiliary (higher-rank) gauge fields
on the world-sheet/world-volume which are absent in the standard
string/brane theories. The main new property of the modified-measure
string/brane models is that the string/brane tension appears as an additional 
dynamical degree of freedom which is canonically conjugated to the auxiliary
world-sheet/world-volume gauge fields. It acquires the physical meaning of 
world-sheet electric field-strength (in the string case) or field-strength of 
higher-rank world-volume gauge fields (in the brane case) obeying the Maxwell
(or Yang-Mills) equations of motion or their higher-rank generalizations.
As a simple consquence of the latter, modified-measure string/brane models
provide (already on the classical level) simple mechanisms for ``color''
charge confinement. Furthermore, in the quantized string context the 
interplay between the intrinsic conformal anomaly and the dynamical nature
of the string tension imply an important constraint on the form of the
dynamical string tension forcing it to be a square of a free massless
world-sheet scalar. It is curious to note that the last property resembles
the property found in the context of the string-inspired low-energy effective
field theory in $D=10$ \ct{siegel} 
where the square of the target-space dilaton field plays the role of a
covariant integration measure density

More detailed study of the effects resulting from the new physical
properties of the modified-measure string models with dynamical tension
reported above will be done in a separate work.

\section*{Acknowledgments} 
Two of us (E.N. and S.P.) are sincerely grateful to the organizers
of the Fifth Workshop on \textsl{Lie Theory and Applications in Physics} 
(Varna, Bulgaria) for the kind invitation to present there the above results. 
E.N. and S.P. also acknowledge cordial hospitality and illuminating discussions
with Prof. Yannis Bakas at the Univesity of Patras (Greece).
The work of E.N. and S.P. is partially supported by NATO collaborative
linkage grant \textsl{PST.CLG.978785} and Bulgarian NSF grant \textsl{F-904/99}. 
Finally, all of us acknowledge support of our collaboration through the exchange
agreement between the Ben-Gurion Univesity of the Negev and the Bulgarian
Academy of Sciences.


\end{document}